\documentclass[3p,times]{elsarticle}
\usepackage{ecrc}
\usepackage{color}

\volume{00}

\firstpage{1}

\journalname{Nuclear Instruments and Methods in Physics Research A}

\runauth{}

\jid{procs}

\jnltitlelogo{\textbf{\footnotesize NUCLEAR INSTRUMENTS \& METHODS IN PHYSICS RESEARCH\\} \scriptsize Section A}

\CopyrightLine{2013}{Published by Elsevier Ltd.}

\usepackage{amssymb}

\usepackage[figuresright]{rotating}

\usepackage{amssymb,bm,mathrsfs,bbm,amscd}
\usepackage[tbtags]{amsmath}
\usepackage{lastpage}
\usepackage{lineno}
\usepackage{subfigure}
\usepackage{epstopdf}
\usepackage{amsmath}
\usepackage{overpic}
\usepackage{footnote}
\usepackage{threeparttable}
\makesavenoteenv{table}
\usepackage[colorlinks,linkcolor=red,anchorcolor=green,citecolor=blue]{hyperref}

\usepackage{multicol}
\usepackage{array,tabularx}
\usepackage{graphicx}
\usepackage{subfigure}
\usepackage{lineno}
\usepackage{ulem}
\begin{document}

\begin{frontmatter}



\dochead{}
\title{{\color{red}}First test of a $\mathrm{CdMoO_4}$ scintillating bolometer for neutrinoless double beta decay experiments with $\mathrm{{}^{116}Cd}$ and $\mathrm{{}^{100}Mo}$ nuclides}
\author{M. Xue\corref{cor1}\fnref{label1,label2}}\cortext[cor1]{Corresponding author.}\ead{xuemx@mail.ustc.edu.cn}
\author{D.V. Poda\fnref{label3,label4}}
\author{Y. Zhang\fnref{label1,label2}} 
\author{H. Khalife\fnref{label3}}
\author{A. Giuliani\fnref{label3,label5}}
\author{H. Peng\fnref{label1,label2}} 
\author{P. de Marcillac\fnref{label3}}
\author{E. Olivieri\fnref{label3}}
\author{S. Wen\fnref{label1,label2}} 
\author{K. Zhao\fnref{label1,label2}} 
\author{Y. Wei\fnref{label1,label2}} 
\author{V. Novati\fnref{label3}}
\author{A.S. Zolotarova\fnref{label3}}
\author{S. Marnieros\fnref{label3}}
\author{C. Nones\fnref{label6}}
\author{T. Redon\fnref{label3}}
\author{Z. Xu\fnref{label1,label2}} 
\author{X. Wang\fnref{label1,label2}} 
\author{P. Chen\fnref{label7}} 
\author{H. Chen\fnref{label7}} 
\author{L. Dumoulin\fnref{label3}}

\address[label1]{State Key Laboratory of Particle Detection and Electronics, University of Science and Technology of China, Hefei 230026, China}
\address[label2]{Department of Modern Physics, University of Science and Technology of China, Hefei 230026, China}

\address[label3]{CSNSM, Univ. Paris-Sud, CNRS/IN2P3, Universit$\acute{e}$ Paris-Saclay, 91405 Orsay, France}
\address[label4]{Institute for Nuclear Research, 03028 Kyiv, Ukraine}
\address[label5]{DISAT, Universit$\grave{a}$ dell'Insubria, 22100 Como, Italy}
\address[label6]{IRFU, CEA, Universit$\acute{e}$ Paris-Saclay, F-91191 Gif-sur-Yvette, France}

\address[label7]{State Key Laboratory Base of Novel Functional Materials and Preparation Science, Key Laboratory of Photoelectric Materials and Devices of Zhejiang Province, Ningbo University, Ningbo 315211, China}

\begin{abstract}
A large cylindrical cadmium molybdate crystal with natural isotopic abundance has been used to fabricate a scintillating bolometer. The measurement was performed above ground at milli-Kelvin temperature, with simultaneous readout of the heat and the scintillation light. The energy resolution as FWHM has achieved from 5 keV (at 238 keV) to 13 keV (at 2615 keV). We present the results of the $\alpha$ versus $\beta$/$\gamma$ events discrimination. The low internal trace contamination of the $\mathrm{CdMoO_4}$ crystal was evaluated as well. The detector performance with preliminary positive indications proves that cadmium molybdate crystal is a promising absorber for neutrinoless double beta decay scintillating bolometric experiments with $\mathrm{{}^{116}Cd}$ and $\mathrm{{}^{100}Mo}$ nuclides in the next-generation technique. 
\end{abstract}

\end{frontmatter}


\section{Introduction}

The study of neutrinoless double beta decay (0$\mathrm{\nu\beta\beta}$) is fruitful for understanding fundamental neutrino properties and has been playing a critical role in particle physics especially after the discovery of neutrino oscillation \cite{0vbb}. Plenty of experiments are now focusing on this interesting research field worldwide \cite{exp}. Observation of the process (Z, A)$\rightarrow$(Z+2, A)+$2e^{-}$ would imply violation of lepton number conservation as well as definitely new physics beyond Standard Model and testify to the Majorana nature of the neutrino and provide an absolute value and hierarchy of the neutrino masses \cite{motivation}.

Cryogenic phonon-scintillation detectors with a heat-light double readout look one of the most promising techniques for searching for this extremely rare decay process \cite{bolom1, bolom2, bolom3, bolom4, bolom5, bolom6, bolom7,bolom8}, owing to their high detection efficiency, good energy resolution, excellent particle discrimination capability by simultaneous measurement of the heat and light channels, and low energy threshold. They also offer the important possibility of using compounds with a choice of different nuclei that are theoretically favorable. 

$\mathrm{CdMoO_4}$ crystal integrated with two interesting target nuclides, $\mathrm{{}^{100}Mo}$  and $\mathrm{{}^{116}Cd}$, would be a promising material to search for 0$\nu\beta\beta$ decay \cite{research}. Both $\mathrm{{}^{100}Mo}$  and $\mathrm{{}^{116}Cd}$ have high transition energy (Table~\ref{tab1}), while the $\gamma$ background mainly ends at 2615 keV, and reasonable natural isotopic abundance (Table~\ref{tab1}). The scintillating characteristics of a $\mathrm{CdMoO_4}$ crystal have been investigated experimentally from 22 K to 300 K in a previous work \cite{research}, especially the fluorescence yields and decay times. Furthermore, the Monte Carlo (MC) study provided convincing evidence that signals of 0$\nu\beta\beta$ of $\mathrm{{}^{116}Cd}$ in the Range Of Interest (ROI) would be higher than the background from the 2$\nu\beta\beta$ events of $\mathrm{{}^{100}Mo}$, which indicates the feasibility and capability of searching for 0$\nu\beta\beta$ process with both $\mathrm{{}^{100}Mo}$  and $\mathrm{{}^{116}Cd}$ nuclides using a heat-scintillation bolometer with $\mathrm{CdMoO_4}$ as a detector-source crystal \cite{research}.

\begin{table}
\centering
\caption{Properties of $\mathrm{{}^{100}Mo}$  and $\mathrm{{}^{116}Cd}$}
\label{tab1}
\begin{tabular}{ccccc}
\hline
Parent isotope& Isotopic abundance (\%) \cite{iso}& Q value (keV) \cite{q}& $T^{2\nu\beta\beta,exp}_{1/2}$ (years) \cite{mo2v,cd}& $T^{0\nu\beta\beta}_{1/2}$ (years) \cite{mo0v,cd} \\
\hline
$\mathrm{{}^{100}Mo}$ & 9.82 & 3034 & (6.90$\pm$0.52)$\times10^{18}$ & $>1.1\times10^{24}$\\
$\mathrm{{}^{116}Cd}$ & 7.49 & 2813 & (2.63$\pm$0.22)$\times10^{19}$ & ${\color{red}}>2.2\times10^{23}$\\
\hline
\end{tabular}
\end{table}

In this work, we report a typical bolometer assembly process with $\mathrm{CdMoO_4}$ crystal at CSNSM (Orsay, France) \cite{mo2v}. We also describe results of an above-ground test of the $\mathrm{CdMoO_4}$ scintillating bolometer at $\sim$25 mK for $\sim$8 h, providing the energy resolution,  $\alpha$/${\gamma}(\beta)$ particle discrimination, quenching factor, and internal radioactive contamination. Finally, in the last part of the paper, we give our conclusions and discuss future prospects. 

\section{Development of the $\mathrm{CdMoO_4}$-based scintillating bolometer}
\begin{figure}[h] 
\center{
\subfigure{
\includegraphics[width=2.0 in, height=1.8 in]{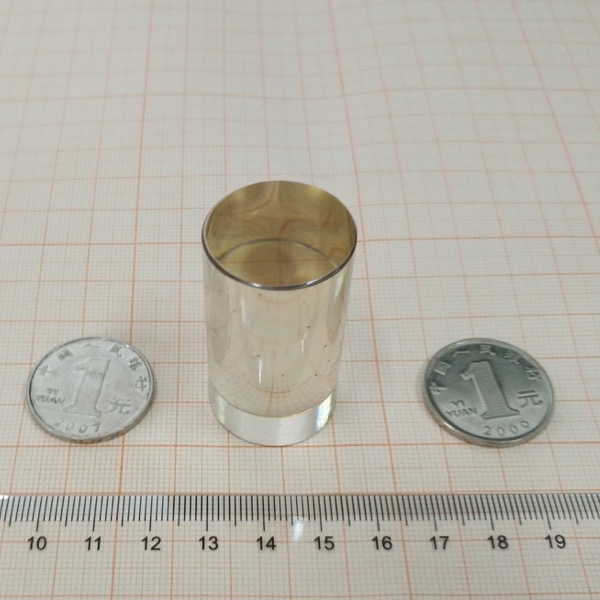}}
\subfigure{
\includegraphics[width=2.0 in, height=1.8 in]{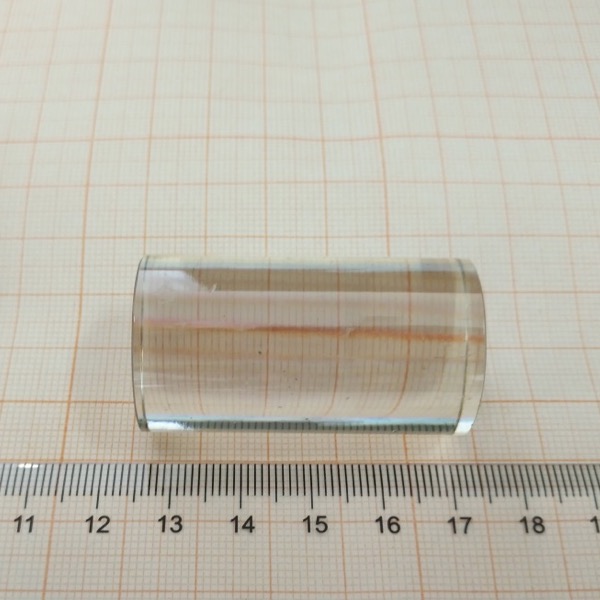}}}
\caption{\label{fig1}Photographs of the $\phi{25\times45} $ mm $\mathrm{CdMoO_4}$ crystal tested as a scintillating bolometer.}
\end{figure}

\begin{figure}[h] 
\center{
\subfigure[]{
\includegraphics[width=2.5 in]{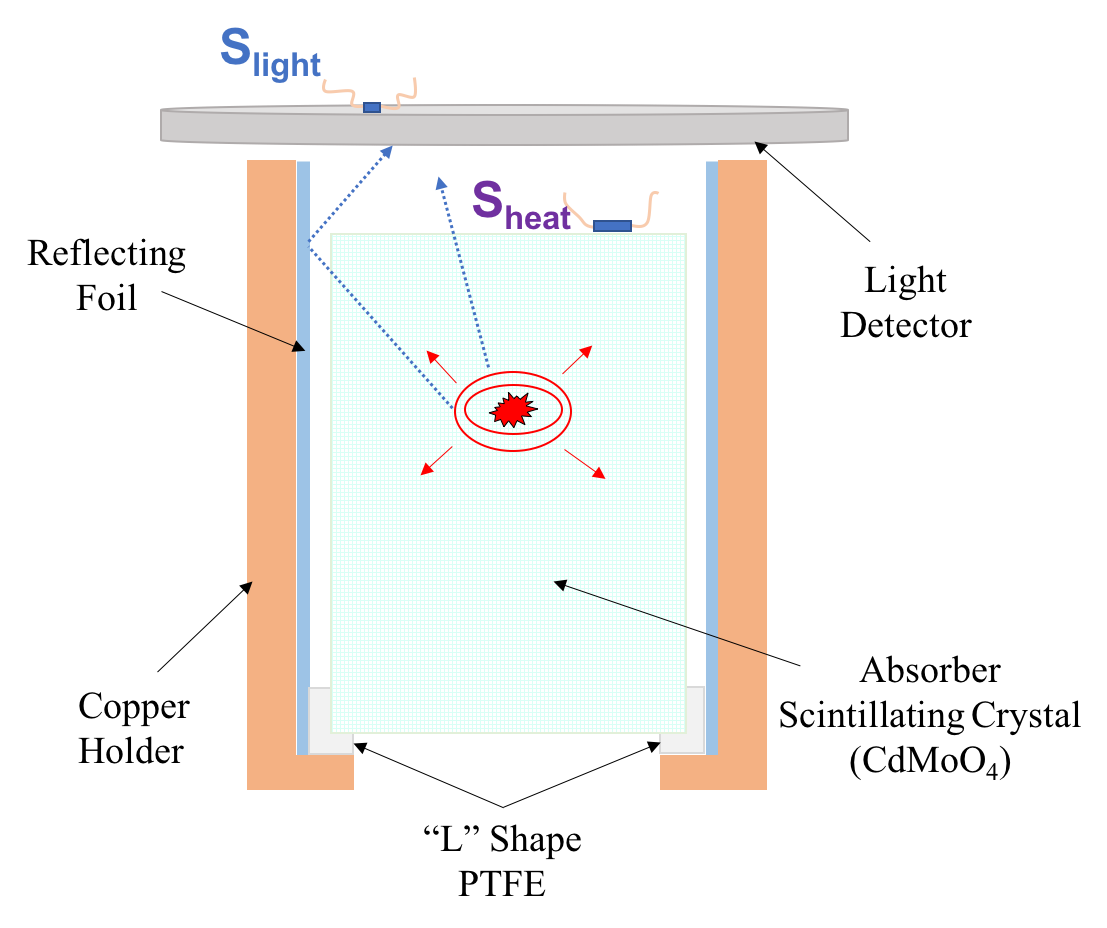}}
\subfigure[]{
\includegraphics[width=2.0 in]{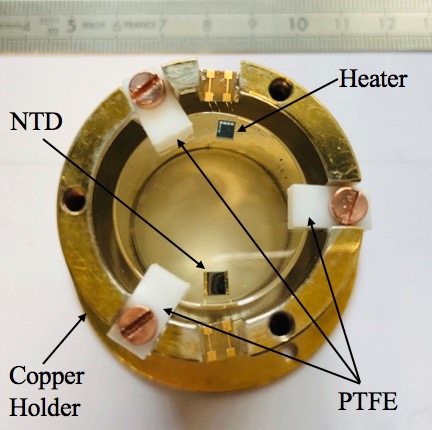}}}
\caption{\label{fig2}The 134.1 g $\mathrm{CdMoO_4}$ scintillating bolometer assembled inside a copper frame covered by reflecting foil. A Kapton pad is used as the electronic connection bridge. PTFE clamps are used as a thermal link to fix the crystal with copper screws. An NTD-Ge sensor and a  silicon heater are glued onto the surface of the $\mathrm{CdMoO_4}$ crystal. (a) A scheme of the detector. (b) Photograph of the $\mathrm{CdMoO_4}$ bolometer.}
\end{figure}

The $\mathrm{CdMoO_4}$ crystal produced at Ningbo Universiity \cite{cmo} has a mass and size of, respectively, 134.1 g and $\phi{25\times45} $ mm. All the surfaces of the crystal are polished to optical grade, as shown in Fig.~\ref{fig1}. The $\mathrm{CdMoO_4}$ scintillating crystal sits inside a copper frame coated with gold, which is of a special design with an extra edge on the bottom. We used PTFE pieces as a thermal link to separate the crystal and the copper holder and fixed the crystal with screws, as displayed in Fig.~\ref{fig2}. The inner surface of the detector holder is covered by reflecting foil to improve the scintillating light collection. On the top, there is a {\color{red}}typical CSNSM fabricated Neganov-Luke-assisted Light Detector (LD) \cite{bolom4}, which is to collect photons emitted from the $\mathrm{CdMoO_4}$ scintillating crystal. Two Neutron Transmutation Doped (NTD) Germanium thermistors are used for both heat and light signal readout. An NTD with a size $3\times3\times1$ mm${}^{3}$ is glued onto the crystal using Araldite \cite{mo2v}; this bicomponent epoxy is radiopurity at the demanded level and it is deposited in spots, thereby mediating the different thermal contractions. For LD, the temperature sensor has approximately three-times-smaller volume with the aim to reduce the heat capacity and to increase its sensitivity \cite{bolom8, ld}. A silicon resistor, glued onto the surface of the $\mathrm{CdMoO_4}$ crystal, was used to produce a calibrated heat pulse in order to monitor the thermal gain of the bolometer and do the stabilization\cite{stab1,stab2}. This is indeed subject to variation upon temperature drifts of the cryostat that can spoil the energy resolution. 

The NTDs and heaters are then wire bonded to the Kapton pads on the edge of the copper holder with a golden wire of diameter 25 $\mu$m. Two wires are used for each electrical connection to provide redundancy. Twisted copper wires are soldered to the pads with lead-free solder.

\section{Detector operation and data analysis}
The low-temperature tests of the $\mathrm{CdMoO_4}$ scintillating bolometer were performed in a cryogenic laboratory at CSNSM by using a pulse-tube cryostat housing a high-power dilution refrigerator \cite{refrigerator}. Because it is an above-ground measurement, shielding made of low radioactivity lead was placed outside the cryostat to reduce the number of the pile-up events from random coincidences of environmentally induced events. The heat and light pulses, produced by a particle interacting in the crystal and transduced in a voltage pulse by the NTD thermistors, are amplified and fed into a data-taking system. The streamed data were recorded by a 16-bit ADC with a 10-kHz sampling frequency. For the 8 h of data used in this context, the operating temperature was 25 mK and there was a radioactive source, $\mathrm{{}^{232}Th}$, outside the cryostat to do energy calibration during the measurement.

\subsection{Energy calibration}
\begin{figure}[h]
\centering
\includegraphics[width=10cm]{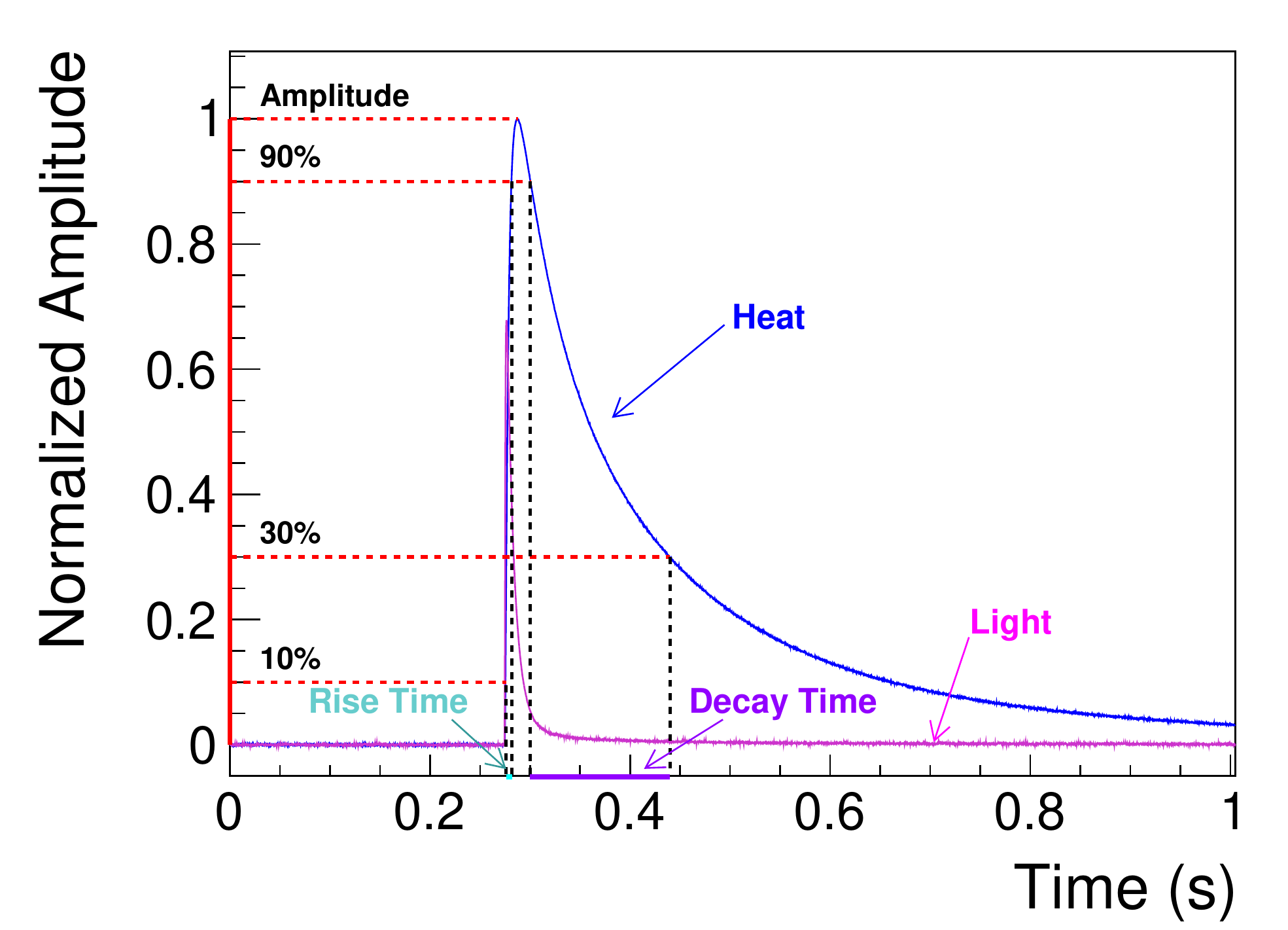}
\caption{\label{fig3}Typical thermal pulses: the heat pulse (blue) with normalized amplitude, rise time and decay time defined and the small light pulse (pink) with a fast time response. All these useful parameters shown here can be used during the whole data processing and analyses, such as amplitude for energy calibration, combined decay time and rise time for Pulse Shape Discrimination (PSD).}
\end{figure}

The collected data were processed offline by CSNSM tools \cite{ana1,ana2}. After applying primary event selection criteria \cite{ana1}, useful information was obtained for each event, including amplitude, baseline, decay time, rise time, and etc., as shown in Fig.~\ref{fig3}.

\begin{figure} 
\centering

\subfigure[]{
\begin{minipage}[t]{0.46\linewidth}
    \centering
    \includegraphics[width=8cm]{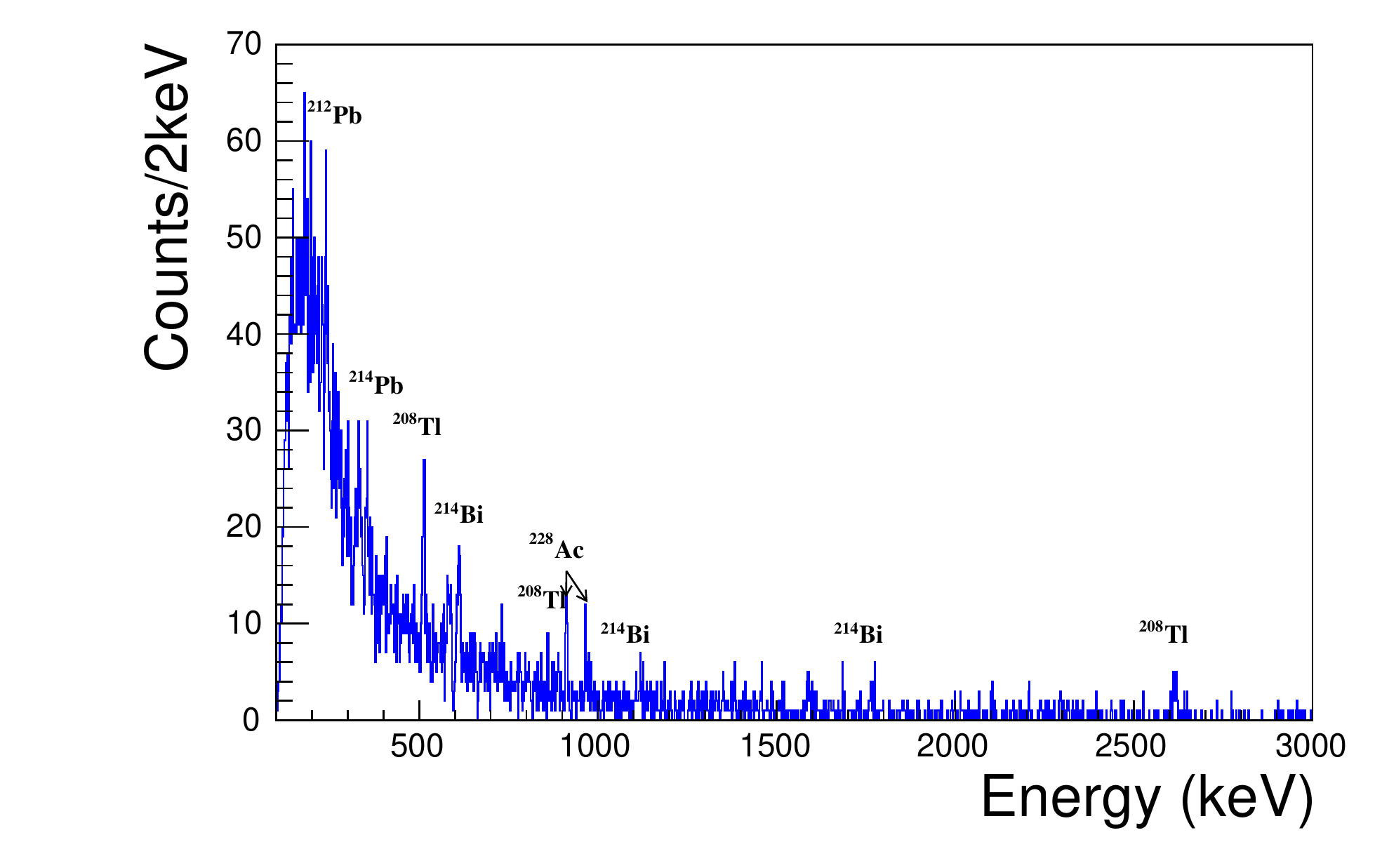}
\end{minipage}
}
\subfigure[]{
\begin{minipage}[t]{0.46\linewidth}
    \centering
    \includegraphics[width=8cm]{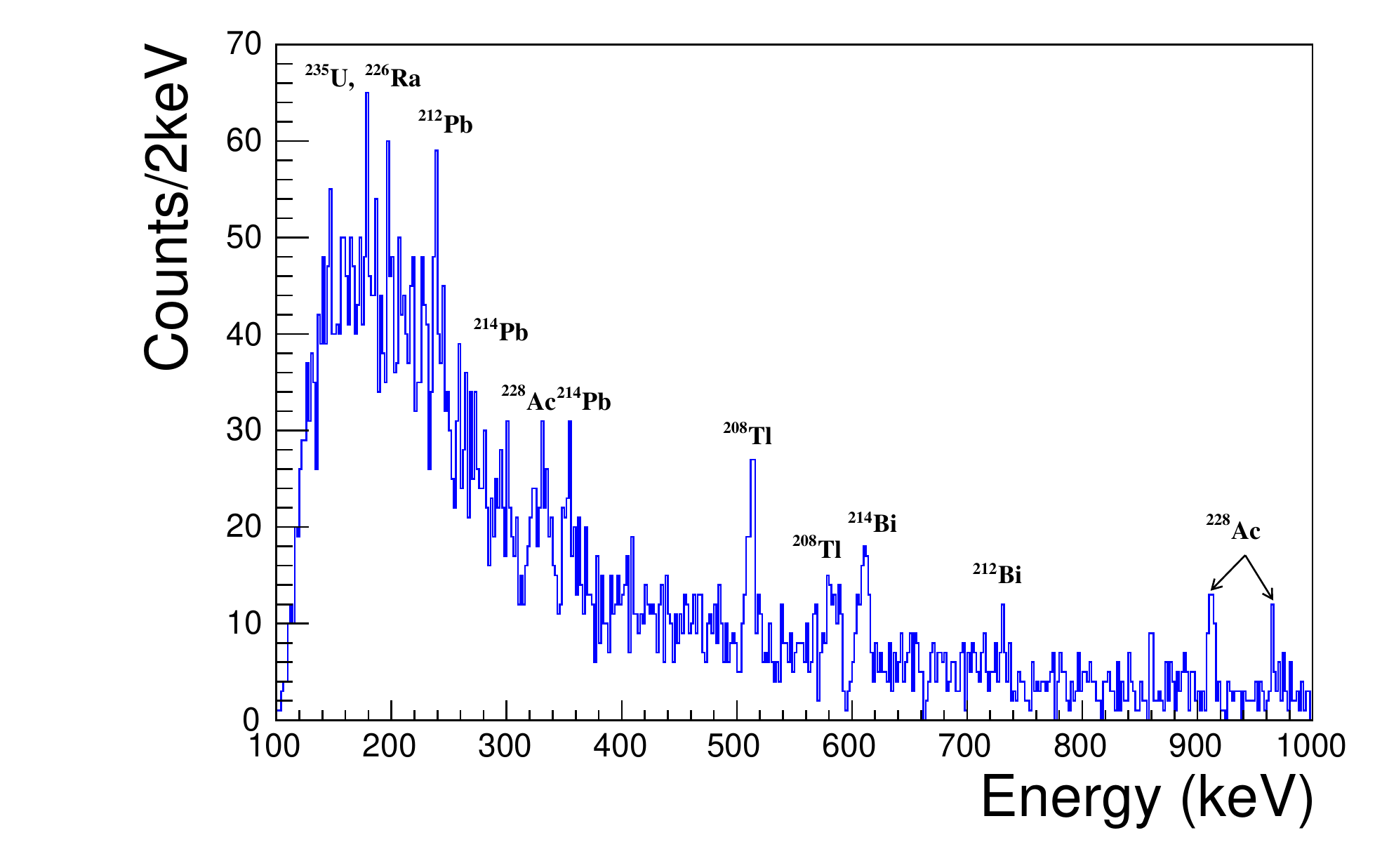}
\end{minipage}
}
\centering
\caption{\label{fig4}Energy spectra obtained with a 134.1 g $\mathrm{CdMoO_4}$ scintillating bolometer in 8 h long calibration measurements performed above ground at CSNSM.}
\end{figure}

In this measurement, the typical $\mathrm{\gamma}$ lines with high intensity of $\mathrm{{}^{212}Pb}$ (238 keV),  $\mathrm{{}^{208}Tl}$ (583 keV),  $\mathrm{{}^{208}Tl}$ (2615 keV), $\mathrm{{}^{214}Pb}$ (351 keV) and $\mathrm{{}^{214}Bi}$ (609 keV) (the last two are actually from the inner contamination of $\mathrm{^{238}U}$ chain) are identified and used for calibration afterwards. The heat response of the scintillating bolometer is calibrated as seen in Fig.~\ref{fig4}. Despite the low statistics, the intrinsic peaks of the nuclides are clearly observable.

For cryogenic detectors used in the 0$\nu\beta\beta$ experiment, the energy resolution should be better than 10 keV Full Width at Half Maximum (FWHM) at the ROI \cite{bolom8, mo2v, resolution} , which is one of the most crucial requirements. {\color{red}}The energy resolution of this $\mathrm{CdMoO_4}$ bolometer can be achieved by fitting the typical $\mathrm{\gamma}$ peaks. In each fit, the signal shape is modeled as a Gaussian function, and the background is described with a linear function. The energy spectrum is performed with an unbinned maximum-likelihood fit to the accepted candidate events. The peaks' FWHM are evaluated as 5$\pm$2 keV at 238 keV and 13$\pm$4 keV at 2615 keV, which show that the energy resolution of the $\mathrm{CdMoO_4}$ crystal detector is acceptable but still needs optimization. Reducing size of the absorber and lowering the testing temperature (25 mK not being typical for these bolometer tests) could possibly improve the energy resolution, as they would be expected to increase the detector sensitivity. Long-time measurement is necessary to increase the statistics and improve the accuracy of the final results.

The LD in this experimental setup was calibrated with cosmic rays. The simultaneous and independent readout of the heat and scintillation light offer the oppotunity to do anticoincidence measurements, which means that we can look at the events that just went through the germanium wafer (LD). GEANT4 simulations were used to model the LD response with isotropic cosmic rays, especially the energy deposition. For anticoincidence cosmic muon events, there is an LD signal only without a heat signal of the crystal in both the experimental data and the GEANT4 simulation data. From the experimental sample, the ADC histogram responding to the energy deposited by the muon can be obtained, and from the MC sample the energy deposited by muon Minimum Ionizing Particles (MIPs) can be evaluated. The energy calibration of the LD has been done as $21.4\pm0.3$ ADC/keV.

\subsection{$\mathrm{\alpha}$ versus $\mathrm{\beta/\gamma}$ discrimination} 
\begin{figure} 
\centering

\subfigure[]{
\begin{minipage}[t]{0.47\linewidth}
    \centering
    \includegraphics[width=8cm]{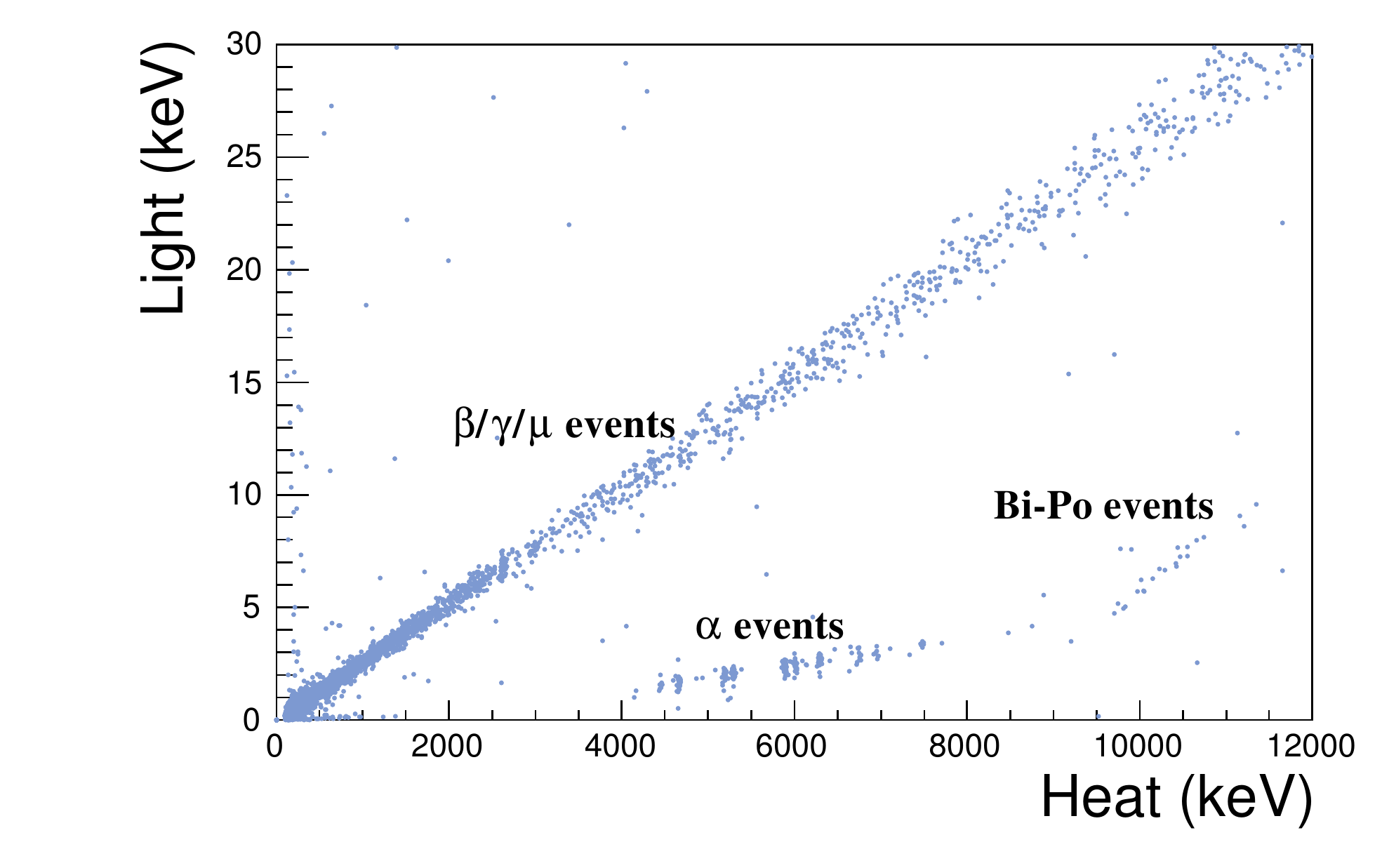}
\end{minipage}
}
\subfigure[]{
\begin{minipage}[t]{0.47\linewidth}
    \centering
    \includegraphics[width=8cm]{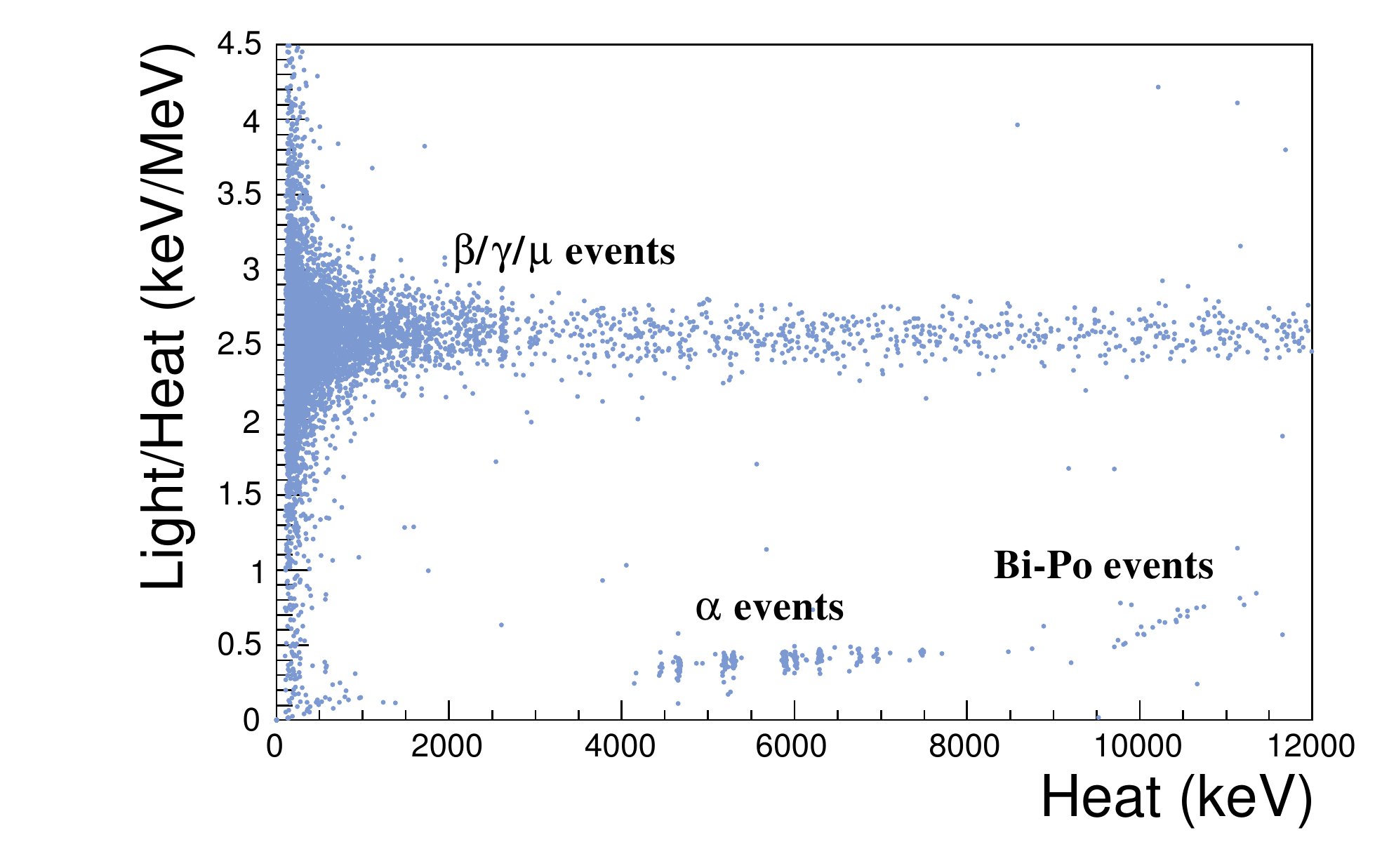}
\end{minipage}
}
\centering
\caption{\label{fig5}(a) The two-dimensional scatter plot obtained from the heat-light double readout of the $\mathrm{CdMoO_4}$ bolometer. It is possible to recognize three main structures. A prominent fully-populated band contains $\mathrm{\beta}$, $\mathrm{\gamma}$, and cosmic muons events. A second clear feature of the scatter-plot is a cluster of points with a much lower light emission with respect to the main band as $\mathrm{\alpha}$ events. A third structure observable in the scatter-plot is the Bi-Po events above 8 MeV. There is a modestly populated band at low energies, with a much lower slope with respect to the main $\mathrm{\beta}$ band, which is nuclear recoil band induced by ambient neutron scattering off nuclei inside the detector. In addition, there is no obvious feature to exhibit the (anti)correlation between the light and heat channels. (b) The light-to-heat energy ratio as a function of the heat energy for the calibration spectrum. $\mathrm{\beta/\gamma}$ and $\mathrm{\alpha}$ decays give rise to very clear separate distributions.}
\end{figure}

The usual way to visualize $\mathrm{\alpha}$ versus $\mathrm{\gamma/\beta}$ discrimination is to draw the light versus heat scatter-plot (Fig.~\ref{fig5}(a)). This graph is built by acquiring in coincidence the heat and light pulses for the same event. Here each event is identified by a point whose abscissa is equal to the heat signal amplitude (recorded by the scintillating bolometer) and whose ordinate is equal to the light signal amplitude (simultaneously recorded by the LD). The scatter-plot contains two main separated populations: a band of $\mathrm{\gamma}$ ($\mathrm{\beta}$ and muons)-induced events and a distribution of $\mathrm{\alpha}$ decays caused by trace impurity of the $\mathrm{CdMoO_4}$ sample. Meanwhile, a clear Bi-Po event ($\mathrm{\alpha\beta}$ {\color{red}}cascade decay) is present in the spectrum owing to the slowness of the thermal detectors, making the two decays appear as one \cite{bipo}.

The commonly used particle identification parameter for scintillating bolometers is the Light Yield (LY), shown in Fig.~\ref{fig5}(b), defined as the ratio between the measured light (in keV) and heat (in MeV) signals. The LY for $\mathrm{\gamma}$ quanta, $\mathrm{\beta}$ particles, and cosmic muons in the energy interval 0.2-2.7 MeV is $\sim$2.55 keV/MeV, which is not corrected for the light collection efficiency. The ratio of the LY parameters for $\mathrm{\alpha}$s and $\mathrm{\gamma(\beta)}$s gives the quenching factor for the scintillation light signals of $\mathrm{\alpha}$ particles: $QF_\alpha=LY_\alpha/LY_{\gamma(\beta)}$. The result for the $\mathrm{CdMoO_4}$ detector is showing 16$\%$ quenching of the light emitted by $\mathrm{\alpha}$ particles with respect to the $\mathrm{\gamma(\beta)}$-induced scintillation. 

\begin{figure} 
\centering

\subfigure[]{
\begin{minipage}[t]{0.45\linewidth}
    \centering
    \includegraphics[width=8cm]{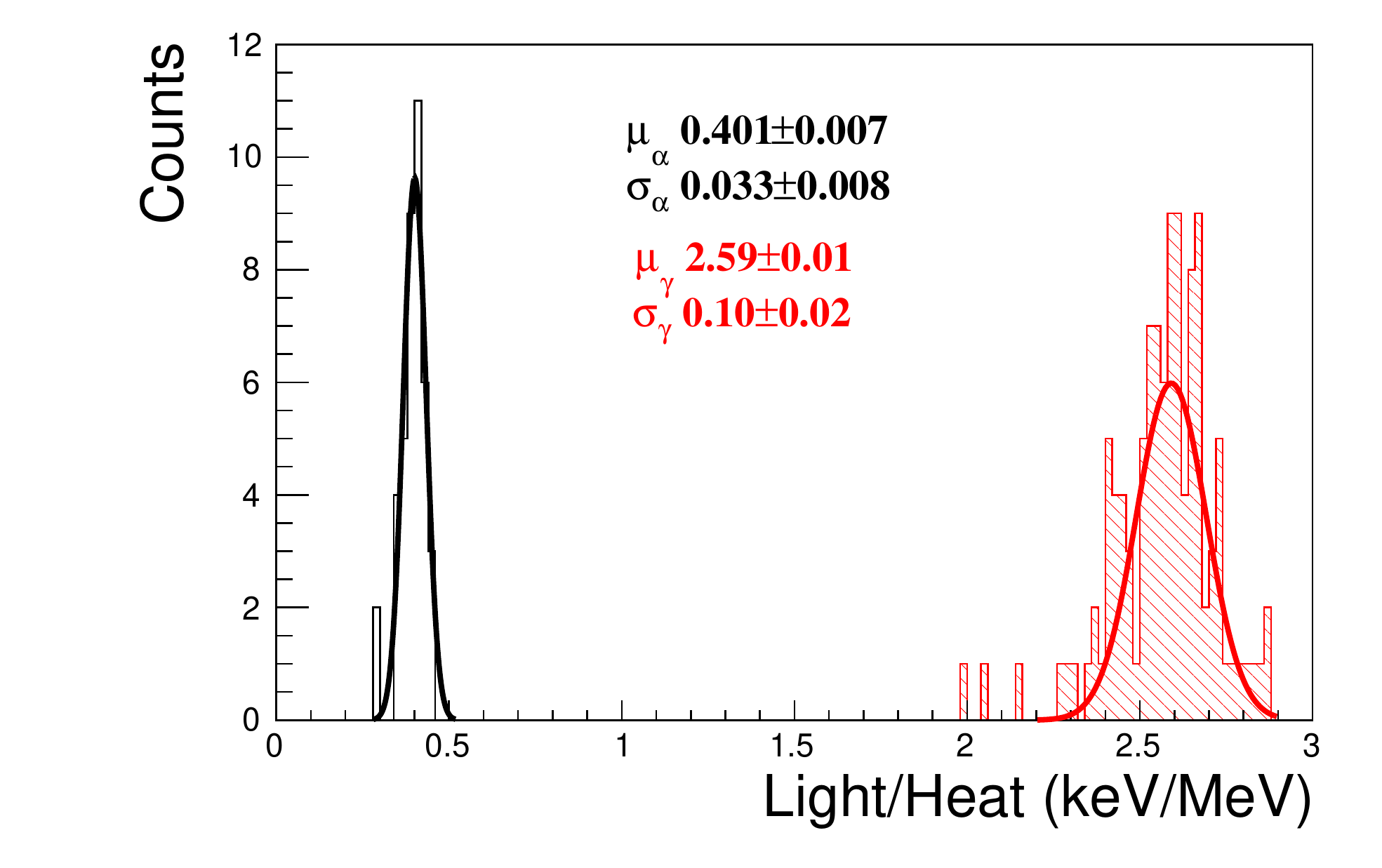}
\end{minipage}
}
\subfigure[]{
\begin{minipage}[t]{0.45\linewidth}
    \centering
    \includegraphics[width=8cm]{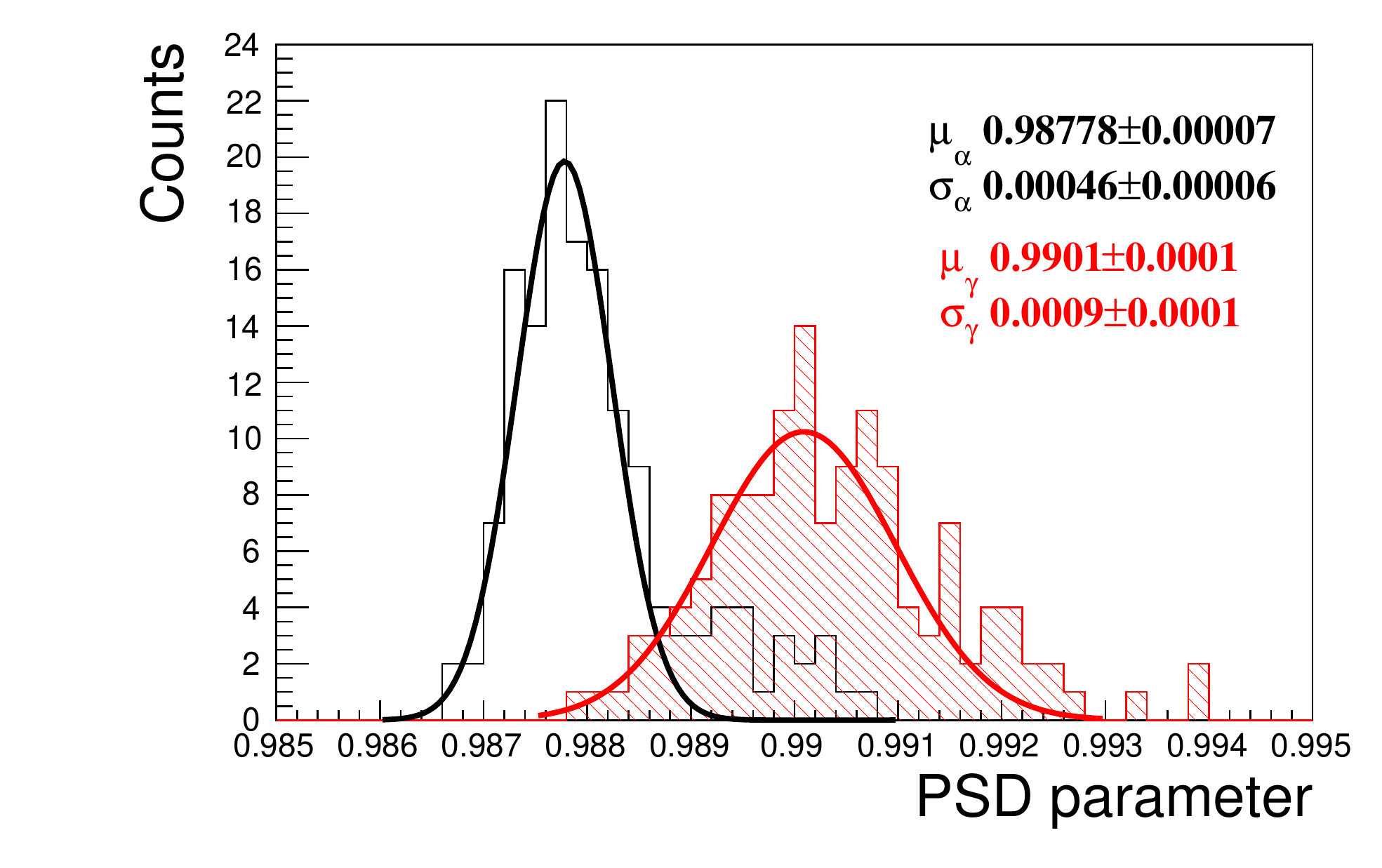}   
\end{minipage}
}
\centering
{\color{red}}\caption{\label{fig6}Evaluation of the $\alpha$ vs. $\gamma/\beta$ discrimination ability in $\mathrm{CdMoO_4}$ bolometer. (a) Combined with light-heat double readout information, $\alpha$ events ($\mathrm{4.7-5.6\ MeV}$) are in black,  $\gamma/\beta$ particles ($\mathrm{2.5-3.5\ MeV}$) are in red. The separation reported as DP is 21. (b) PSD parameter histogram of the events for $\mathrm{4.0-6.5\ MeV}$ $\alpha$ samples (black) and $\mathrm{2.5-3.5\ MeV}$ $\gamma/\beta$ events (red), which just use the heat signal to exhibit the pulse-shape difference between $\alpha$ particles and $\gamma/\beta$'s. The discrimination parameters are obtained by Gaussian fits, the achieved DP value is 2.}
\end{figure}

{\color{red}}In particular, we can also show the particle discriminating ability of the bolometer by defining a typical DP (Discrimination Power) value \cite{dp}:

\begin{equation}
DP=\frac{\mu_{\gamma/\beta}-\mu_{\alpha}}{\sqrt{\sigma^{2}_{\gamma/\beta}+\sigma^{2}_{\alpha}}}
\end{equation}
In view of the fact that the $\mathrm{CdMoO_4}$ bolometer is a light-heat double readout system, we can adopt the histogram shown in  Fig.~\ref{fig6}(a), which is obtained by projecting the plot of  Fig.~\ref{fig5}(b) onto the vertical axis, and respectively fitting with Gaussian functions to acquire the corresponding parameters. In that case, the achieved $\alpha/\gamma(\beta)$ DP=21. The premise is to extend the $\alpha$ region ($\mathrm{4.7-5.6\ MeV}$) backward to the $\gamma(\beta)$ energy region ($\mathrm{2.5-3.5\ MeV}$). {\color{red}}In addition, we can use the PSD technology as well to read out information only by using the heat signal without considering the light signal, according to the difference in the readout pulse of the heat signal from different particle types. As shown in  Fig.~\ref{fig6}(b), the particles are identified relying on the pulse shape information of the $\alpha/\gamma(\beta)$ particles response spectrum. All the parameters of interest are obtained by fitting Gaussian functions respectively to obtain a DP value of 2. 

\subsection{$\mathrm{\alpha}$ contamination}
\begin{figure}
\centering
\includegraphics[width=10cm]{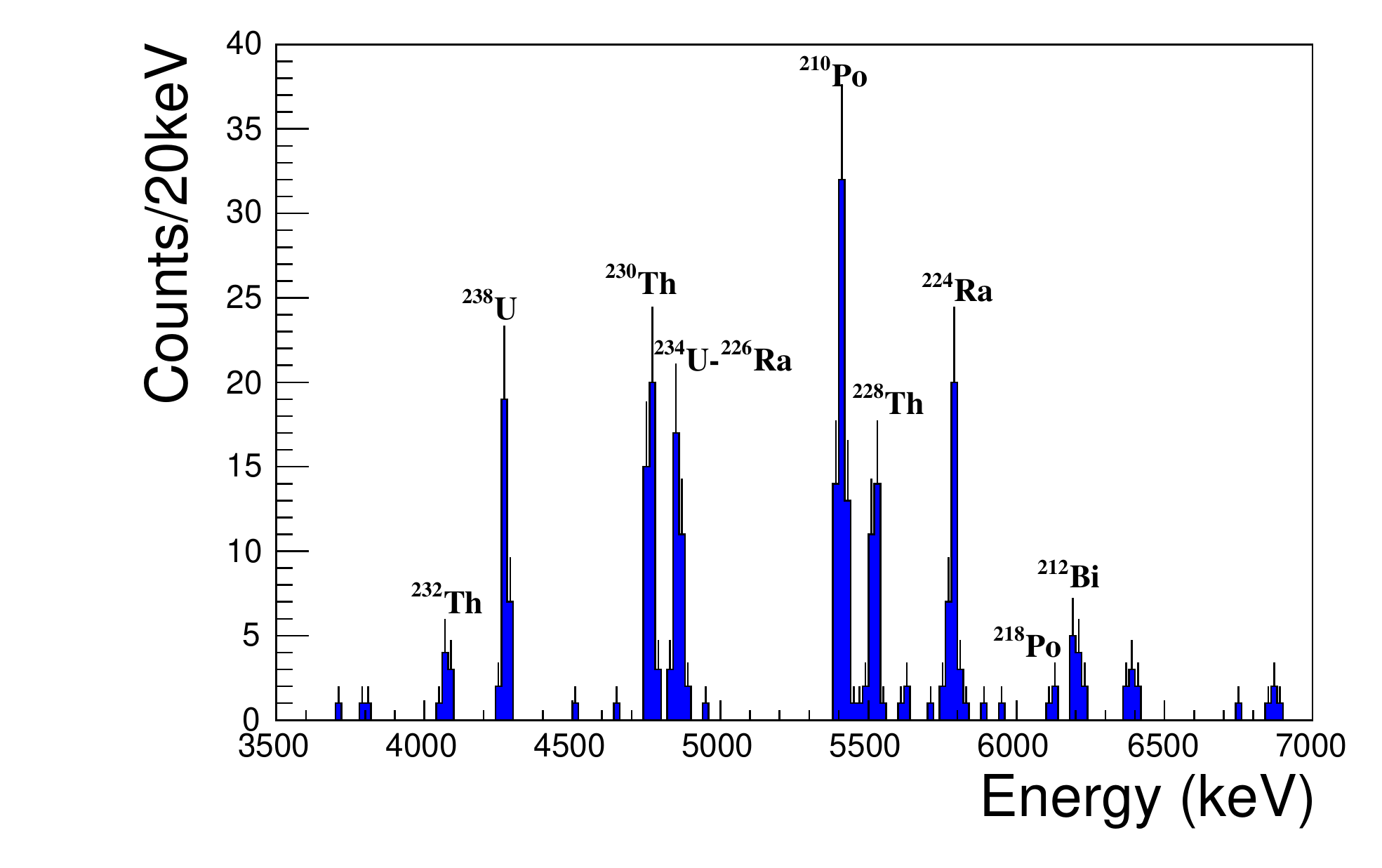}
\caption{\label{fig7}Background spectrum of $\mathrm{\alpha}$ events accumulated by the natural $\mathrm{CdMoO_4}$ crystal with 8 h of data taking. The spectrum was energy-calibrated using $\mathrm{{}^{238}U}$ (4269 keV), $\mathrm{{}^{234}U}$ (4859 keV), and $\mathrm{{}^{210}Po}$ (5407 keV), the three most prominent $\mathrm{\alpha}$ peaks. The $\mathrm{\alpha}$-induced backgrounds show clear peaks, which we can identify in U/Th chains.}
\end{figure}

As pointed out in the previous section, the internal contaminations were evaluated in the same run by using the energy spectrum of the $\mathrm{\alpha}$ events. The events were selected under the associated LY to be below 1 keV/MeV. Because of possible thermal quenching of the heat signals of alpha particles, the energy calibration using $\mathrm{\gamma}$ may not be suitable for $\mathrm{alpha}$ events. Therefore, to present the correct energy of the $\mathrm{\alpha}$ events, an additional calibration based on the identification of $\mathrm{\alpha}$ peaks is needed (see Fig.~\ref{fig7}). The peak of $\mathrm{{}^{210}Po}$ was identified in the data; it confirms a broken equilibrium in the radioactive chain \cite{bipo,alpha1, alpha2}, because, in secular equilibrium, the activity for all radioactive sources should share the same value. $\mathrm{{}^{238}U}$ (and its daughters $\mathrm{{}^{234}U}$, $\mathrm{{}^{230}Th}$, $\mathrm{{}^{226}Ra}$, and $\mathrm{{}^{210}Po}$ events) and $\mathrm{{}^{228}Th}$ (with daughter $\mathrm{{}^{224}Ra}$) were detected in the $\mathrm{CdMoO_4}$ crystal. We use the number counting method to estimate the radioactive contamination in the $\mathrm{CdMoO_4}$ crystal scintillator, as reported in Table~\ref{tab2}, which have already considered the efficiencies.

\begin{table}
\centering
\begin{threeparttable}
\caption{Radioactive contamination details of the $\mathrm{CdMoO_4}$ crystal (with mass 134.1 g) tested as a scintillating bolometer in above-ground conditions for 8 h of measurement.}
\label{tab2}
\begin{tabular}{ccc}
\hline
Isotopes & $\mathrm{Q_\alpha}$\tnote{1} keV&  Activity(mBq/kg)\\
\hline
$\mathrm{{}^{238}U}$ & 4269& 10 $\mathrm{\pm}$ 3 \\ 
$\mathrm{{}^{230}Th}$ & 4769& 14 $\mathrm{\pm}$ 4\\
$\mathrm{{}^{234}U}$, $\mathrm{{}^{226}Ra}$ & 4859/4870 & 12 $\mathrm{\pm}$ 3 \\
$\mathrm{{}^{210}Po}$ & 5407& 22 $\mathrm{\pm}$ 5\\

$\mathrm{{}^{232}Th}$ & 4081& 3 $\mathrm{\pm}$ 2 \\
$\mathrm{{}^{228}Th}$ & 5520& 11 $\mathrm{\pm}$ 3 \\
$\mathrm{{}^{224}Ra}$ & 5788&  12 $\mathrm{\pm}$ 3\\

\hline
\end{tabular}

\begin{tablenotes}
\footnotesize
\item[1] Both the $\mathrm{\alpha}$ and recoil nuclear kinetic energy are contained in the bolometer
\end{tablenotes}
\end{threeparttable}
\end{table}

There is clearly some improvement that can be made regarding the radiopurity level of $\mathrm{CdMoO_4}$ crystal. The original material (powder) of the crystal was not with the requirement of radiopurity. Moreover, a strong segregation of the radioactive impurities in the $\mathrm{CdMoO_4}$ crystal growing process is also the first emergency. It is crucial to control the quality of the crystals.

\section{Conclusions}
We successfully tested a 134.1 g cylindrical $\mathrm{CdMoO_4}$ crystal (with a size of $\mathrm{\phi25\times45}$ mm and a natural abundance) for about 8 h as a scintillating bolometer, exploiting heat-light dual readout at 25 mK in an above-ground cryostat and demonstrating the feasibility of such a technique. Calibrations with different typical $\mathrm{\gamma}$ lines were performed by {\color{red}}radioactive source $\mathrm{{}^{232}Th}$, put outside the cryostat. The $\mathrm{CdMoO_4}$ detector exhibits a high energy resolution with FWHM from 5 keV to 13 keV in the energy range 0.2-2.6 MeV. The measurements demonstrated excellent detector performance in terms of energy resolution and $\mathrm{\alpha}$ versus $\mathrm{\beta/\gamma}$  separation power. Thanks to the simultaneous readout of heat and light, the bolometer shows a discrimination power as 21 in the 2.5-5.6 MeV region. Moreover, since there is no any kind of {\color{red}}raw material selection, the observed radioactive contamination in U/Th chains from $\mathrm{CdMoO_4}$ crystal is in a level of a few mBq/kg. Taking into account the experimental temperature condition (25 mK not being optimum) and the quality of the crystal (without any requirement of radiopurity), we conclude that there are some improvements that can be made. An R$\&$D of $\mathrm{CdMoO_4}$ crystal scintillating bolometers with the aim of optimization is in progress, especially the crystal purification, the ongoing negotiation of underground measurement in China Jinping Underground Laboratory or others. We believe that the reported results are very encouraging in view of a next generation of neutrinoless double beta decay experiments based on this technique.

\section{Acknowledgements}
{\color{red}}The authors Mingxuan Xue, Yunlong Zhang, Haiping Peng, Sicheng Wen, Kangkang Zhao, Yifeng Wei, Zizong Xu and Xiaolian Wang thank the strong supports from the major projects of Double First-Class University of University of Science and Technology of China. The works is supported by the National Natural Science Foundation of China under Project No. 11625523. The PhD fellowship of H. Khalife has been partially funded by the P2IO LabEx (ANR-10-LABX-0038) managed by the Agence Nationale de la Recherche (France) in the framework of the 2017 P2IO Doctoral call.




\bibliographystyle{elsarticle-num}

\end{document}